\documentclass[twocolumn,a4paper,prd]{revtex4}

\usepackage{bm}
\usepackage{epsfig}
\usepackage{graphics}
\usepackage{natbib}
\usepackage{fancyh}
\usepackage[l]{floatflt}
\usepackage{epsfig}
\usepackage{amssymb}
\usepackage{latexsym}
\usepackage{times}
\usepackage{slashed}
\usepackage{upgreek}
\usepackage{amsmath}
\usepackage{amsfonts}
\usepackage{amsbsy}
\usepackage{amscd}
\usepackage{bbm}

\newcommand{\Eref}[1]{Eq.~(\ref{#1})}

\newcommand{\Sref}[1]{Sec.~\ref{#1}}
\newcommand{\Fref}[1]{Fig.~\ref{#1}}
\newcommand{\Tref}[1]{Table~\ref{#1}}
\newcommand{\cref}[1]{Ref.~\cite{#1}}

\newcommand{\hepth}[1]{{\ftn \tt hep-th/#1}}
\newcommand{\hepph}[1]{{\ftn\tt hep-ph/#1}}

\newcommand{\arxiv}[1]{{\ftn\tt  arXiv:#1}}

\newcommand{\bal}{\begin{align}}
\newcommand{\eall}{\end{align}}
\newcommand{\beqs}{\begin{subequations}}
\newcommand{\eeqs}{\end{subequations}}
\newcommand{\eec}{\end{center}}
\newcommand{\bec}{\begin{center}}

\newcommand{\eem}{\end{matrix}}
\newcommand{\bem}{\begin{matrix}}
\newcommand{\eeq}{\end{equation}}
\newcommand{\beq}{\begin{equation}}
\newcommand{\ba}{\begin{array}}
\newcommand{\ea}{\end{array}}
\newcommand{\bea}{\begin{eqnarray}}
\newcommand{\eea}{\end{eqnarray}}
\newcommand{\baq}{\begin{eqnarray}}
\newcommand{\eaq}{\end{eqnarray}}
\newcommand{\bel}{\begin{align}}
\newcommand{\eel}{\end{align}}

\newcommand\eqs[2]{Eqs.~(\ref{#1}) and (\ref{#2})}

\newcommand\eqss[3]{Eqs.~(\ref{#1}), (\ref{#2}) and (\ref{#3})}

\newcommand{\ftn}{\footnotesize}

\newcommand{\TeV}{{\mbox{\rm TeV}}}

\newcommand{\GeV}{{\mbox{\rm GeV}}}
\newcommand{\EeV}{{\mbox{\rm EeV}}}
\newcommand{\PeV}{{\mbox{\rm PeV}}}

\newcommand{\meV}{{\mbox{\rm meV}}}
\newcommand{\keV}{{\mbox{\rm keV}}}

\newcommand{\etal}{{\it et al.\/}}

\def\lf{\left(}
\def\rg{\right)}
\newcommand\vev[1]{\left\langle {#1} \right\rangle}

\newcommand{\Gr}{\ensuremath{\widetilde{G}}}

\newcommand{\Vhi}{\ensuremath{V_{\rm I}}}

\newcommand{\Hhi}{\ensuremath{H_{\rm I}}}

\newcommand{\Ns}{\ensuremath{{N_\star}}}

\newcommand{\mP}{\ensuremath{m_{\rm P}}}

\newcommand{\la}{\ensuremath{\lambda_{\rm a}}}

\def\openone{\leavevmode\hbox{\small1\kern-3.8pt\normalsize1}}
\newcommand{\dV}{\ensuremath{\Delta V_{\rm I}}}

\newcommand{\fns}{\ensuremath{f_{N\star}}}
\newcommand{\fnf}{\ensuremath{f_{N{\rm f}}}}
\newcommand{\fn}{\ensuremath{f_{N}}}
\newcommand{\pn}{\ensuremath{p_{N}}}

\newcommand{\Ggr}{\ensuremath{{\Gamma}_{3/2}}}

\newcommand{\Gsn}{\ensuremath{{\Gamma}_{\dzv}}}
\newcommand{\Gth}{\ensuremath{{\Gamma}_{\th}}}
\newcommand{\Gh}{\ensuremath{{\Gamma}_{\tilde{h}}}}

\newcommand{\msn}{\ensuremath{m_{z}}}

\newcommand{\hd}{{\ensuremath{H_d}}}
\newcommand{\hu}{{\ensuremath{H_u}}}

\newcommand{\ks}{\ensuremath{k_\star}}
\newcommand{\ns}{\ensuremath{n_{\rm s}}}
\newcommand{\as}{\ensuremath{a_{\rm s}}}
\newcommand{\As}{\ensuremath{A_{\rm s}}}

\newcommand{\eph}{\ensuremath{\epsilon}}

\newcommand{\ep}{\ensuremath{\epsilon}}

\newcommand{\Ve}{\ensuremath{V}}

\newcommand{\Vs}{\ensuremath{V_{\rm I\star}}}
\newcommand{\Hs}{\ensuremath{H_{\rm I\star}}}

\newcommand{\zv}{\ensuremath{Z_{\rm v}}}

\newcommand{\nnp}{\ensuremath{n_+}}
\newcommand{\nnm}{\ensuremath{n_-}}
\newcommand{\nnpm}{\ensuremath{n_\pm}}

\newcommand{\what}{\ensuremath{\widehat}}
\newcommand{\wtilde}{\ensuremath{\widetilde}}
\newcommand{\mss}{\ensuremath{\widetilde m}}

\newcommand{\vcc}{\ensuremath{V_{\Lambda}}}

\newcommand{\vrm}{\ensuremath{{\rm v}}}

\newcommand{\vo}{\ensuremath{v_{0}}}
\newcommand{\va}{\ensuremath{v_{1}}}
\newcommand{\vb}{\ensuremath{v_{2}}}
\newcommand{\vc}{\ensuremath{v_{3}}}
\newcommand{\jo}{\ensuremath{J_{0}}}

\newcommand{\Om}{\ensuremath{\Omega}}
\newcommand{\om}{\ensuremath{\omega}}
\newcommand{\omz}{\ensuremath{\Omega_{,Z}}}
\newcommand{\omzz}{\ensuremath{\Omega_{,ZZ^*}}}

\def\bz{{Z^*}}

\def\al{{\alpha}}

\def\bt{{\beta}}

\def\n{\bar{n}}

\def\th{{\theta}}

\newcommand{\Trh}{\ensuremath{T_{\rm rh}}}
\newcommand{\sg}{\ensuremath{z}}

\newcommand{\sgo}{\ensuremath{z_0}}
\newcommand{\ko}{\ensuremath{k_0}}
\newcommand{\sgx}{\ensuremath{z_\star}}
\newcommand{\sgf}{\ensuremath{z_{\rm f}}}

\newcommand{\dz}{\ensuremath{\delta z}}
\newcommand{\dzx}{\ensuremath{\delta z_{\star}}}
\newcommand{\dzf}{\ensuremath{\delta z_{\rm f}}}
\newcommand{\dk}{\ensuremath{\delta k}}
\newcommand{\dzv}{\ensuremath{{\updelta} z}}
\newcommand{\dzvh}{\ensuremath{\what{\updelta z}}}

\newcommand{\Ld}{\ensuremath{\Lambda}}

\newcommand{\se}{\ensuremath{\widehat z}}
\newcommand{\sex}{\ensuremath{\widehat{z}_\star}}
\newcommand{\sef}{\ensuremath{\widehat{z}_{\rm f}}}

\newcommand{\mgr}{\ensuremath{m_{3/2}}}

\newcommand{\mz}{\ensuremath{ m_{z}}}
\newcommand{\mth}{\ensuremath{m_{\theta}}}

\newcommand{\cm}{\ensuremath{C_{\Lambda}}}

\newcommand{\Ctp}{\ensuremath{C_{\om}^+}}
\newcommand{\Ctm}{\ensuremath{C_{\om}^-}}
\newcommand{\Ctpm}{\ensuremath{C_{\om}^\pm}}

\newcommand{\phc}{\ensuremath{\Phi}}

\def\Ka{K\"{a}hler potential}
\def\Kmn{K\"{a}hler manifold}
\def\Kaa{K\"{a}hler~}

\newcommand{\plk}{{\it Planck}}

\renewcommand{\refname}{{\bf\scshape References}}

\renewcommand{\thesubsection}{{\small\sf\Alph{subsection}}}

\renewenvironment{subequations}{%
\refstepcounter{equation}%
\setcounter{parentequation}{\value{equation}}%
  \setcounter{equation}{0}
  \ignorespaces
}{%
  \setcounter{equation}{\value{parentequation}}%
  \ignorespacesafterend
}

  \makeatletter
\renewcommand*\l@section[2]
  {%
    \ifnum \c@tocdepth >\z@
      \addpenalty \@secpenalty
      \addvspace {1.em \@plus \p@ }%
      \setlength \@tempdima {1.8em}%
      \begingroup
        \parindent \z@
        \rightskip \@pnumwidth
        \parfillskip -\@pnumwidth
        \leavevmode \bfseries
        \advance \leftskip \@tempdima
        \hskip -\leftskip
        #1\nobreak
        \hfil
        \nobreak
        \hb@xt@ \@pnumwidth {\hss #2\kern -\p@ \kern \p@ }%
        \par
      \endgroup
    \fi
  }
\makeatother

\begin{document}

\title{\bf\scshape  Inflection-Point Sgoldstino Inflation in no-Scale Supergravity}

\author{{\scshape Constantinos Pallis} \\ {\small\it Laboratory of Physics, Faculty of
Engineering, Aristotle University of Thessaloniki, GR-541 24
Thessaloniki, GREECE} \\ {\ftn\sl  e-mail address: }{\ftn\tt
kpallis@gen.auth.gr}}

\begin{abstract}

\noindent {\ftn \bf\scshape Abstract:} We propose a modification
of no-scale supergravity models which incorporates sgoldstino
stabilization and supersymmetry (SUSY) breaking with a tunable
cosmological constant by introducing a K\"ahler potential which
yields a kinetic pole of order one. The resulting scalar potential
may develop an inflection point, close to which an inflationary
period can be realized for subplanckian field values consistently
with the observational data. For central value of the spectral
index $\ns$, the necessary tuning is of the order of $10^{-6}$,
the tensor-to-scalar ratio $r$ is tiny whereas the running of
$\ns$, $\as$, is around $-3\cdot10^{-3}$. Our proposal is
compatible with high-scale SUSY and the results of LHC on the
Higgs boson mass.
\\\\  {\scriptsize {\sf PACs numbers: 12.60.Jv, 04.65.+e}
\hfill {\sl\bfseries Published in} {\sl Phys. Lett. B} {\bf 843},
138018 (2023)}

\end{abstract}\pagestyle{fancyplain}

\maketitle

\rhead[\fancyplain{}{ \bf \thepage}]{\fancyplain{}{\sl
Inflection-Point Sgoldstino Inflation in no-Scale SUGRA}}
\lhead[\fancyplain{}{\sl C. Pallis}]{\fancyplain{}{\bf \thepage}}
\cfoot{}

\vspace*{-1.5cm}
\section{\bfseries\scshape Introduction}\label{intro}\vspace*{-.3cm}

One of the most tantalizingly evasive problems in Particle Physics
is the explanation of the large hierarchies existing in the
fundamental scales of the modern theories. Two of these scales are
related to the cosmological problems of inflation and \emph{Dark
Energy} ({\ftn\sf DE}) whereas a third one is related to the scale
of \emph{Supersymmetry} ({\ftn\sf SUSY}) breaking. The first of
the scales above, is expected to be less than $10^4~\EeV$, the
second one is related to the present acceleration of the Universe
and is really very tiny ($\Lambda\sim 1~\meV$) whereas the scale
of the SUSY partners, $\mss$, is continuously pushed more and more
beyond the $\TeV$ region due to the lack of any positive signal at
LHC until now.

In a set of recent papers \cite{ns89, de}, two of the
aforementioned scales ($\mss$ and $\Ld$) are systematically
interconnected within the framework of no-scale
\emph{Supergravity} ({\ftn\sf SUGRA}) \cite{old, nilles}. Adopting
simple \Ka s parameterizing flat or curved (compact or
non-compact) \Kmn s, one can initially derive superpotentials
which yield SUSY breaking along Minkowski flat directions.
Combining two types of these superpotentials, \emph{de Sitter}
({\ftn\sf dS}) vacua are achieved which may, in addition, explain
the notorious DE problem finely tuning a single superpotential
parameter. In other words, these models offer a \emph{technically
natural} resolution to the DE problem. The construction can be
easily extended to multi-modular settings of mixed geometry. Mild
deformations of the adopted moduli geometry can cure possible
instabilities and/or massless excitations. \emph{Soft
SUSY-Breaking} ({\sf\ftn SSB}) parameters of the order of the
gravitino ($\Gr$) mass $\mgr$ can be also derived. It is worth
mentioning that the method above does not require any external
mechanism for vacuum uplifting -- see, e.g., \cref{kallosh,
antst}.

It would be certainly beneficial if we could incorporate in the
aforementioned framework observationally acceptable inflation as
done in \cref{nsinfl,nsreview}. There the SUSY breaking sector is
supplemented with the inflationary one which assures
Starobinsky-like inflation consistently with data -- for similar
attempts see \cref{king, lhclinde, ketov, kai, scroest}. In
contrast to those models, though, we here adopt the most
economical possible interface of both sectors postulating that the
scalar component of the goldstino superfield, -- which is
responsible for the SUSY breaking -- plays the role of the
inflaton  -- cf. \cref{postma, ipisugra, ant1, aattr, froest}.
Such a construction predicts directly the scale $\mss$, since the
fundamental mass parameter, $m$, entering the superpotential of
goldstino is relied on the normalization \cite{plcp} of the power
spectrum of the curvature perturbation -- cf.
\cref{ant1,aattr,ipisugra}. As a consequence, $\mgr$ and $\mss$
lie at the $\EeV$ mass scale. Therefore, the set-up of high-scale
SUSY \cite{hall,strumia} naturally arises, as can be demonstrated
by coupling the goldstino sector to the \emph{Minimal SUSY
Standard Model} ({\ftn\sf MSSM}) \cite{martin} and deriving the
relevant SSB parameters \cite{soft}. It is worth mentioning that
these $\mss$ values are compatible with the Higgs boson mass
discovered at LHC \cite{lhc} stabilizing, thereby, the electroweak
vacuum.

To implement the inflationary scenario above we adopt a \Ka\ which
generates a kinetic pole of order one \cite{epole} in the SUGRA
lagrangian. The presence of this pole essentially restricts the
dynamics of all the ingredients of our set-up (inflation, SUSY
breaking and DE) to field values below the Planck scale. For
specific values of the parameters the SUGRA potential develops an
inflection point \cite{ipisusy, ipilinde, aterm,
ipirad,drees,ipisugra} which supports inflationary solutions at
the cost of a tuning of the order of $10^{-6}$ in the selection of
a relevant parameter. The realized \emph{Inflection-Point
inflation} ({\ftn\sf IPI}) is a typical inflationary model -- see,
e.g., \cref{review} -- which was first introduced in the context
of MSSM \cite{ipisusy, aterm} and then employed in various
frameworks \cite{ipilinde, ipirad,drees,ipisugra}. To our
knowledge, none of them combines IPI with the aforementioned
merits of no-scale SUGRA. IPI discussed here occurs for
subplanckian field values and is consistent with data
\cite{plin,bk15} predicting negligible production of primordial
gravitational waves and testable \cite{asdrees} running of $\ns$.
After its end, the universe is reheated up to a temperature
\cite{rh} of $\PeV$ level via the perturbative decay of the heavy
sgoldstino into gravitinos and/or MSSM (s)particles \cite{baerh,
antrh, nsrh, full} via SUGRA-based interactions.

We start our presentation implementing the transition from
Minkowski to dS vacua in the framework of our model in
\Sref{sec1}. Then, in \Sref{sec2}, we verify the generation of the
DE potential energy at the stable dS vacuum  and, in \Sref{sec3},
we extract the SSB terms. In \Sref{sec4} we focus on the
exploration of the inflationary stage and, finally, we summarize
our results in \Sref{con}. Unless otherwise stated, we use units
where the reduced Planck scale $\mP=2.4\cdot 10^{18}~\GeV$ is
taken to be unity, a subscript of type $,\chi$ denotes derivation
\emph{with respect to} ({\small\sf w.r.t.}) the field $\chi$ and
charge conjugation is denoted by a star. We also recall that
$1~\PeV=10^6~\GeV$ and $1~\EeV=10^9~\GeV$.


\section{\bfseries\scshape Model Set-up}\label{sec1}

We work in the context of SUGRA employing just one chiral
superfield $Z$ -- cf. \cref{scroest, froest, ipisugra}. The
F--term SUGRA potential is given by
\beq \Ve=e^{G}\lf G^{ZZ^*} G_Z G_{Z^*}-3\rg,\label{Vsugra} \eeq
where $G$ is the \Kaa-invariant function defined in terms of the
\Ka\  $K=K(Z,Z^*)$ and the superpotential $W=W(Z)$ as follows
\beq \label{Kzz} G := K + \ln |W|^2~~\mbox{with}~~G_{ZZ^*}
=K_{ZZ^*}=K_{,ZZ^*} \eeq
denoting the \Kaa\ metric and $G^{ZZ^*}$ is its inverse. We
concentrate on the following $K$
\beqs\beq \label{Km} K=-N\ln\Omega~~\mbox{with}~~\Omega=
1-(Z+Z^*)/2+k^2\zv^4,\eeq
where $N>0$ and we include in the argument of $\ln$ the
stabilization term
\beq \label{Zv} \zv=Z+Z^*-2\vrm. \eeq\eeqs
Here $k$ and $\vrm$ are two real free parameters. The $Z$ space
generated by $K$ for $k=0$ is hyperbolic and invariant under a set
of transformations related to the group $U(1,1)$ -- see
\cref{epole}. Small $k$'s are completely natural, according to the
't Hooft argument \cite{symm}, thanks to the enhanced symmetry
above.

Following our strategy in \cref{ns89,de} we first substitute $K$
in \Eref{Km} with $k=0$ together with an unspecified $W$ into
\Eref{Vsugra} taking the limit $Z=\bz$ -- the stability of this
path is checked a posteriori below. We obtain
\beq V=\om^{-N}W^2\lf\frac{4\om^2}{N}\lf
\frac{N}{2\om}+\frac{dW}{dZW}\rg^2-3\rg,\label{Vc0}\eeq
where we introduce the shorthand notation
\beq \om=\Om(Z=Z^*,k=0)=1-Z.\label{om}\eeq
A $Z$-flat direction with Minkowski vacua can be assured if we
seek $W=W_0$ such that $V=0$ for any $Z$. $W_0$ can be determined
solving the resulting ordinary differential equation
\beq \frac{d W_{0}}{W_{0}}=\nnpm
\frac{dZ}{\om}~~\mbox{with}~~\nnpm=\frac12\lf
N\pm\sqrt{3N}\rg.\label{Wde}\eeq
For $N=3$ we obtain $\nnpm=0$ and we reveal the archetypal
no-scale model \cite{old} which exclusively yields Minkowski
vacua. For $N\neq3$, we obtain two possible forms of $W_{0}$,
\beq W_0^{(\pm)}(Z)=m\om^{\nnpm}, \label{W0}\eeq
where $m$ is an arbitrary mass parameter which agrees with the one
introduced for the Pol\'onyi model \cite{nilles}. The exponents
$\nnpm\neq0$ may, in principle, acquire any real value, if we
consider $W_0$ as an effective superpotential including
perturbative and non-perturbative contributions from string theory
\cite{ns89}. However, when $N/3$ is a perfect square, integer
$\nnpm$ values may arise too. E.g., for $N=12, 27$ and $48$ we
obtain $(\nnm,\nnp)=(3,9), (9,18)$ and $(18,30)$ respectively.

The solutions in \Eref{W0} can be combined as follows
\beq W_{\Lambda}= W_{0}^{(+)}-\cm W_{0}^{(-)}
=m\om^{\nnp}\Ctm,\label{Wlt}\eeq
where we normalize somehow the relevant coefficients -- cf.
\cref{de} -- setting that multiplying $W_{0}^{(+)}$ equal to
unity. Also,
\beq \Ctpm:=1\pm\cm\om^{-\sqrt{3N}}\label{ctpm}\eeq
and reduces to unity for tiny $\cm$ -- see below. The resulting
potential, $V=V_{\Ld}$, obtained after replacing \eqs{Km}{Wlt}
into \Eref{Vsugra}, is
\beq \vcc=m^2\Om^{-N}\om^{2\nnp} \lf|U/2\om|^2-3|\Ctm|^2\rg,
\label{Vde}\eeq
where we define the quantity
\beq
U=\frac{\sqrt{2N}}{J\Om}\lf\lf\sqrt{3}\Ctp+\sqrt{N}\Ctm\rg\Om+2\sqrt{N}\Ctm\omz\om\rg.
\label{Udef}\eeq
Here $J$ is related to the canonical normalization of the complex
scalar field $Z=ze^{i\th}$ according to which
\beq  \label{Jg} {d\widehat z/
dz}=\sqrt{2K_{ZZ^*}}=J~~\mbox{and}~~ \what{\th}= J\th\sg,\eeq
where $J$ can be expressed in terms of $\Om$ as follows
\beq
J=\sqrt{2N}\lf\frac{\omz^2}{\Om^2}-\frac{\omzz}{\Om}\rg^{1/2}.
\label{jdef}\eeq
The involved derivatives of $\Om$ are found to be
\beq \omz=
-1/2+4k^2\zv^3~~\mbox{and}~~\omzz=12k^2\zv^2.\label{omz}\eeq

The structure of $\vcc$ in \Eref{Vde} is shown in \Fref{fig1}
where the dimensionless quantity $10^2\vcc/m^2\mP^2$ is plotted as
a function of $z$ and $\th$ for the $N, \vrm$, $k$, $m$ and
$C_\Ld$ values shown in \Tref{tab}-- we select the lower $N$ which
yields integer $\nnpm>0$. We see that $\vcc$ develops along the
stable direction $\th=0$ {\sf\ftn (i)} A dS vacuum for
$z=\vrm=0.5$ which interprets DE -- see \Sref{sec2} -- and breaks
SUSY -- see \Sref{sec3} --, and {\sf\ftn (ii)} an inflection point
for $z=\sgo\simeq0.71$, suitable for driving IPI -- see
\Sref{sec4}. The last finding seems to be a consequence of the
adopted $\Om$ in \Eref{Km}, since replacing its $k$-independent
part with $Z+Z^*$ or $1-|Z|^2$ and repeating the procedure above,
the resulting $\vcc$ does not posses inflection point but it
acquires a shape resembling that presented in \cref{ant1, aattr}.

\section{\bfseries\scshape Dark Energy}\label{sec2}

\begin{figure}[t]%
\epsfig{file=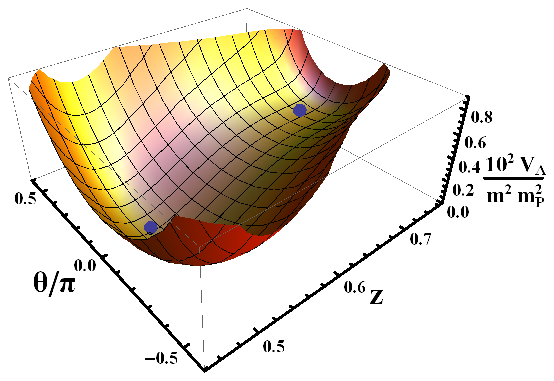,width=8.7cm,,angle=-0}
\vspace*{2.3in} \caption{\sl \small The (dimensionless) SUGRA
potential $10^2\vcc/m^2\mP^2$ in \Eref{Vde} as a function of $z$
and $\th$ defined above \Eref{Jg} for the inputs shown in
\Tref{tab}. The location of the dS vacuum at
$(\vev{z},\th)=(0.5,0)$ and the inflection point at
$(\sgo,\th)\simeq(0.71,0)$ is also depicted by two thick 
points.}\label{fig1}
\end{figure}

The stability of $\vcc$ at its dS -- for $\cm>0$ -- vacuum noticed
by \Fref{fig1} can be analytically verified in general. Namely, we
can show that the vacuum
\beq
\vev{z}=\vrm~~\mbox{and}~~\vev{\th}=0~~\mbox{with}~~\vev{\vcc}=12\cm
m^2 \label{vcc}\eeq
is stable against fluctuations of the various excitations for
$N>3$. In fact, the resulting masses squared of the canonically
normalized scalars are found to be
\beqs\bea \label{mt}
m_{z}&=&48\mgr kN^{-1/2}\vev{\om^{3/2}{\Ctp}/{\Ctm}};\\
m_{\th}&=&2\mgr\lf1-(3/N)\vev{\Ctp/\Ctm}^2\rg^{1/2}\label{mbt}\eea\eeqs
and the $N>3$ assures $m_{\th}^2>0$. The \Gr\ mass contained in
the expressions above can be determined as follows
\beq \label{mgr}
\mgr=\vev{e^{G/2}}=\vev{m\om^{\sqrt{3N}/2}\Ctm}.\eeq
Since $\om<1$ from \Eref{om} and $\sqrt{3N}/2>1$, we infer that
$\mgr\leq m$ and so all the aforementioned masses share
approximately the same size. This is also confirmed from the
sample values accumulated in the third from the bottom row of
\Tref{tab}.

An interpretation of DE can be achieved by demanding
\beq \label{omde} \vev{\vcc}=\Omega_\Lambda\rho_{\rm
c0}=7.2\cdot10^{-121}\mP^4,\eeq
where $\Omega_\Lambda=0.6889$ and $\rho_{\rm
c0}=2.4\cdot10^{-120}h^2\mP^4$ with $h=0.6732$ \cite{plcp} is the
density parameter of DE and the current critical energy density of
the universe. As shown in \Tref{tab}, the required value of $\cm$
signals a serious fine tuning whose the origin remains elusive
within our proposal. However, this value does not influence the
remaining sectors of the model and can be selected in the
definition of $W_{\Lambda}$.

\begin{table}[t!]
\caption{\sl Sample-Point Parameters With Units Reinstalled \\
(Values in curly brackets are obtained by the analytic
expressions)}
\begin{ruledtabular}
\begin{tabular}{c|c|c|c|c}
\multicolumn{5}{c}{\sc Model Parameters}\\\colrule
$m/\mP$&$\cm/10^{-108}$&$N$&$k/0.1$&$\vrm/\mP$\\\colrule
$5.6\cdot10^{-7}$&$2.5$&$12$&$4.0167291$&$0.5$\\\colrule
\multicolumn{5}{c}{$(\nnp,\nnm)=(9,3)$}\\\colrule
\multicolumn{5}{c}{\sc Inflection Point Localization}\\\colrule
$\ko/0.1$&$\sgo/0.1\mP$&$\dk/10^{-6}$&\multicolumn{2}{c}{$\dzx/10^{-4}\mP$}\\\colrule
$4.0166971$&$7.07433$&$3.20232$&\multicolumn{2}{c}{$-1.5$~\{$-1.1$\}}\\\colrule
\multicolumn{5}{c}{\sc Expansion Parameters}\\\colrule
$\vo/(m\mP)^2$&$\va/(m\mP)^2$&$\vb/(m\mP)^2$&$\vc/(m\mP)^2$&$\jo$\\\colrule
$3.9\cdot10^{-3}$&$1.5\cdot10^{-6}$&$-2.1\cdot10^{-6}$&$2.2$&$5.4$\\\colrule
\multicolumn{5}{c}{\sc Inflation Results}\\\colrule
$\ns$&$r/10^{-8}$&$-\as/10^{-3}$&$10^5\As^{1/2}$&$\Ns$\\\colrule
$0.966$~\{$0.97$\}&$4.8$~\{$3.9$\}&$3.3$~\{$3.2$\}&$4.59$~\{$4.27$\}&$46.5$~\{$45$\}\\\colrule
$\Vs^{1/4}/\EeV$&$\Hs/\EeV$&\multicolumn{2}{c|}{$\dzf/10^{-2}\mP$}&$m_{\th\rm
I\star}/\Hs$\\\colrule
$4.6\cdot10^{5}$&$49.5$&\multicolumn{2}{c|}{$-1.16$~\{$-0.87$\}}&$5.1$\\\colrule
\multicolumn{5}{c}{\sc Post-Inflation Results}\\\colrule
$\msn/\EeV$&$\mth/\EeV$&$\mgr/\EeV$&$\Ggr/\keV$&$\Trh/~\PeV$\\\colrule
$319$&$281$&$162$&$85$&$4.9$\\\colrule
\multicolumn{1}{c||}{$\what\mu/\EeV$}&$\mss$&$|A|/\EeV$&$|B|/\EeV$&$|M_{\rm
a}|/\EeV$\\\colrule
\multicolumn{1}{c||}{$81$}&$162$&$1024$&$1200$&$81.1$
\end{tabular}\label{tab}
\end{ruledtabular}
\end{table}


\section{\bfseries\scshape SUSY Breaking}\label{sec3}

The SUSY breaking occurred at the vacuum in \Eref{vcc} can be
transmitted to the visible world if we specify a reference low
energy model. We here adopt MSSM and the total superpotential,
$W_{\rm \Ld O}$, of the theory takes the form \cite{soft}
\beq \label{Who} W_{\rm \Ld O}=W_\Ld(Z) + W_{\rm
MSSM}\lf\phc_\al\rg,\eeq
where $W_{\rm MSSM}$ has the well-known form written in short as
\beq W_{\rm MSSM}=h_{\al\bt\gamma} \phc_\al\phc_\bt\phc_\gamma/6
+\mu\hu\hd\label{wo}\eeq
with the various chiral superfields encoded as
\bea\phc_\al= {Q}, {L}, {d}^c, {u}^c, {e}^c,
\hd~~\mbox{and}~~\hu,\nonumber \eea
and we suppress the generation indices. We also denote the three
non-vanishing Yukawa coupling constants as $h_{\al\bt\gamma}=h_D,
h_U$ and $h_E$ for $(\al,\bt,\gamma)=(Q,\hd,d^c), (Q,\hu,u^c)$ and
$(L,\hd,e^c)$ respectively. As we see below, our model fits well
with the high-scale SUSY \cite{hall,strumia} and therefore $\mu$
acquires values close to $\mgr$. We here handle it as a free
parameter. On the other hand, we consider two simple variants of
the total $K$ of the theory, $K_{\rm  \Ld O}$, ensuring SSB
parameters for $\phc_\al$:
\beqs\bel
K_{1\rm\Ld O}&=K(Z)+\mbox{$\sum_\al$}|\phc_\al|^2;~~~\label{K1}\\
K_{2\rm\Ld O}&=K(Z)+N_{\rm
O}\ln\left(1+\mbox{$\sum_\al$}|\phc_\al|^2/N_{\rm
O}\right),\label{K2} \end{align}\eeqs
where $N_{\rm O}$ may remain unspecified. Note that if we expand
$K_{\rm 2\Ld O}$ for low $\phc_\al$ values, the result
coincidences with $K_{\rm 1\Ld O}$.

Adapting the general formulae of \cref{soft}, we find universal
(i.e., $\wtilde m_\al=\mss$ and $A_{\al\bt\gamma}=A$) SSB terms in
the effective low energy potential which can be written as
\beq V_{\rm SSB}=\wtilde m^2 |\phc_\al|^2+\lf\frac16 A\what
h_{\al\bt\gamma} \phc_\al\phc_\bt\phc_\gamma+ B\what\mu
H_uH_d+{\rm h.c.}\rg \label{vmssm} \eeq
where the normalized (hatted) parameters are defined as
\beq  (\what h_{\al\bt\gamma},\what\mu)=\vev{\om}^{-N/2}(
h_{\al\bt\gamma},\mu),\label{norm} \eeq
whereas the SSB parameters are found to be
\beq\label{mAB} \wtilde m=\mgr,~~|A|=\sqrt{3N}\mgr~~\mbox{and}~~
|B|=(1+\sqrt{3N})\mgr.\eeq
Values for these parameters are displayed in \Tref{tab} for
$\what\mu=\mgr/2$. We see that $|A|>\mgr$ and $|B|>\mgr$  due to
large $N$ adopted there. However, these parameters have very
suppressed impact on the SUSY mass spectra.

Similar values for the gauginos of MSSM are also expected. E.g.,
we may select the gauge-kinetic function \cite{soft} as
\beq f_{\rm a}= \la Z\, \label{gkf} \eeq
where $\la$  is a free parameter absorbed by a redefinition of the
relevant spinors and ${\rm a}=1,2,3$ runs over the factors of the
gauge group of MSSM, $U(1)_Y$, $SU(2)_{\rm L}$ and $SU(3)_{\rm c}$
respectively. In a such case, we find \cite{soft}
\beq |M_{\rm a}|= \sqrt{3/N}\vev{\om/z}\mgr,\label{Ma} \eeq
which is obviously of the order of $\mgr$ -- see e.g. \Tref{tab}.


Scenarios with large $\mss$, although not directly accessible at
the LHC, can be probed via the measured value of the Higgs boson
mass. Within high-scale SUSY, updated analysis requires
\cite{lhc,strumia}
\beq 3\lesssim\mss/\EeV\lesssim300,\label{highb} \eeq
for degenerate sparticle spectrum, low $\tan\beta$ values and
minimal stop mixing. From \Eref{mAB} and the values in \Tref{tab}
we conclude that our setting is comfortably compatible with the
requirement above.

\section{\bfseries\scshape Inflection-Point Inflation}\label{sec4}

We analyze here the inflationary sector of our model. In
\Sref{sec41} we outline our method for the determination of the
inflection point of the potential and in \Sref{sec42} we describe
our semi-analytic approach to inflationary dynamics. Then, in
\Sref{sec43}, we discuss the reheat process and in \Sref{sec44}
the parameters of our model are confronted with observations.

\begin{figure}[!t]\vspace*{-.25in}\centering
\includegraphics[width=60mm,angle=-90]{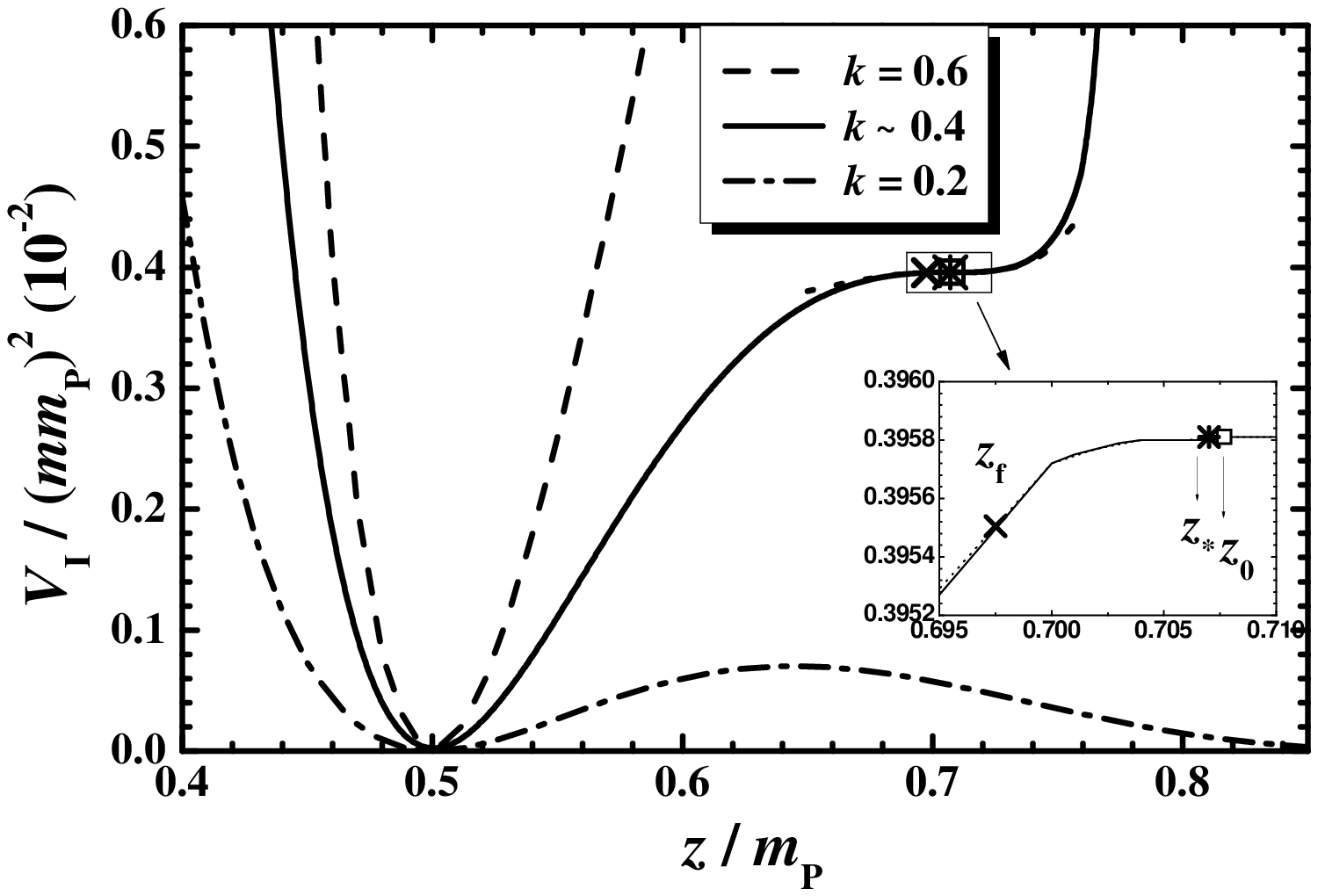}
\caption{\sl \small Dimensionless inflationary potential
$\Vhi/m^2\mP^2$ as a function of $\sg$ for $k=0.40167291$ (black
line) or $k=0.2$ (dot-dashed line) or $k=0.6$ (dashed line) and
the remaining inputs in \Tref{tab}. The dotted line is obtained by
\Eref{vappr}. The values of $\sgx$, $\sgf$ and $\sgo$ (for the
first case) are also indicated.} \label{fig2}
\end{figure}

\subsection{\sc\small\sffamily Inflection-Point Conditions}\label{sec41}

The inflationary potential $\Vhi=\Vhi(z)$ is obtained from
$\vcc(Z)$ in \Eref{Vde}, setting $\th=0$ and $\cm\simeq0$.  Let us
initially clarify that $\Vhi$ develops discontinues due to the
denominator of $U$ in \Eref{Udef} or the numerator of $J$ in
\Eref{jdef}. In view of \Eref{omz}, these singularities can be
determined by solving numerically the equation
\beq
256k^4(z-\vrm)^6+16k^2(z-\vrm)^2(z+2\vrm-3)+1/4=0.\label{poles}\eeq
In \Fref{fig2} we represent the segments of $\Vhi$ as a function
of $z$ which are continuously connected with the vacuum in
\Eref{vcc}. For the model parameters shown in \Tref{tab} $\Vhi$
exhibits poles at the points $(z_{1\rm p},z_{2\rm p},z_{3\rm
p})=(0.26,0.78,1.2)$ and is plotted by a solid line for $z_{1\rm
p}\lesssim z\lesssim z_{2\rm p}$. Thanks to the interplay of the
two opposing contributions in the parenthesis in \Eref{Vde}, a
step is generated for $\vrm<z<z_{2\rm p}$. The emergence of the
inflationary plateau, though, crucially depends on $k$. Indeed,
for $k=0.2$ and $k=0.6$ and keeping the residual inputs in
\Tref{tab}, we obtain the dot-dashed and dashed lines respectively
in \Fref{fig2} where no inflection point is localized. We find
$(z_{1\rm p},z_{2\rm p})=(0.05,1.21)$ in the former case and
$(z_{1\rm p},z_{2\rm p},z_{3\rm p})=(0.34,0.68,1.1)$ in the
latter.

\begin{figure}[!t]\vspace*{-.25in}\centering
\includegraphics[width=60mm,angle=-90]{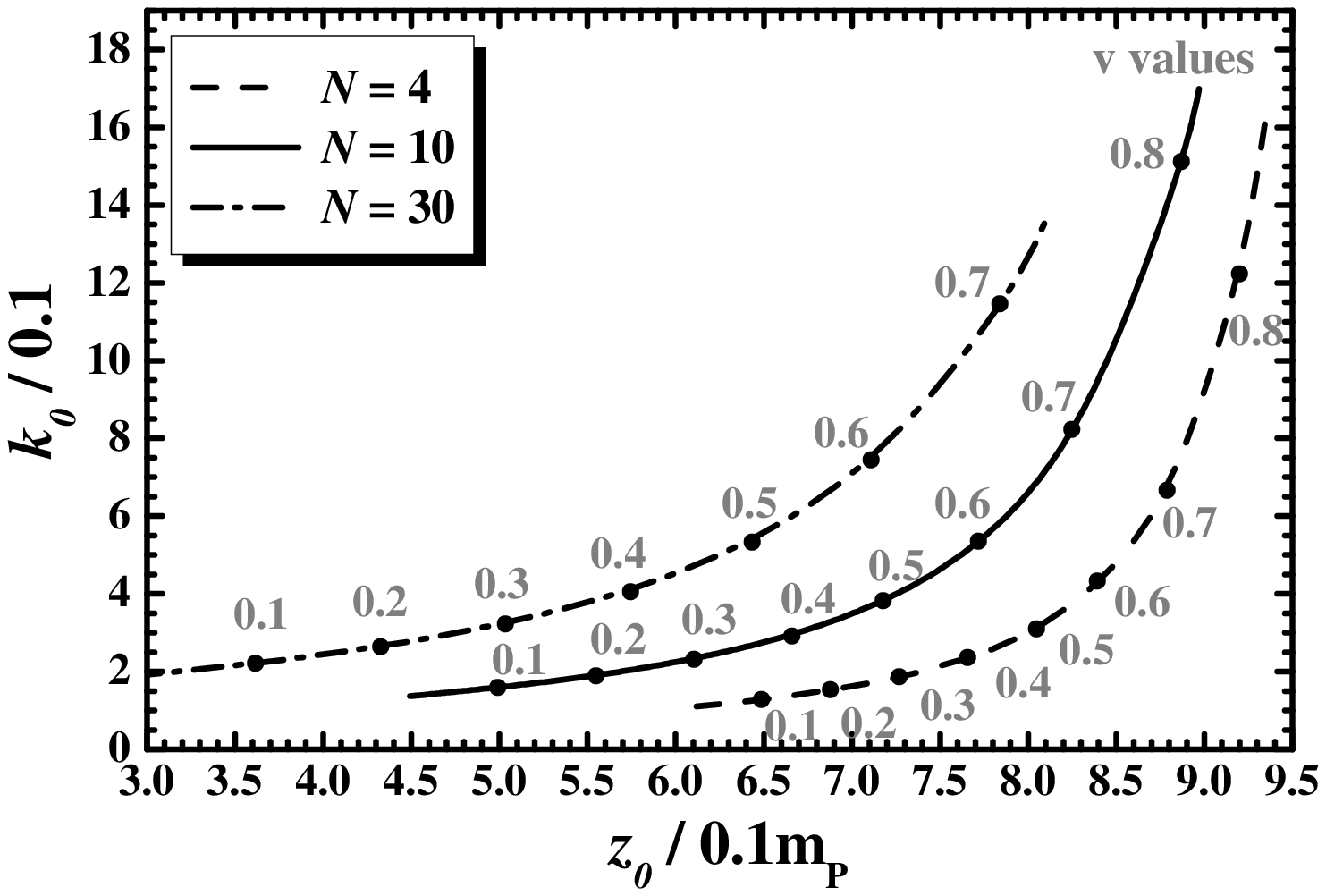}
\caption[]{\sl\small Location of the inflection point in the
$\sgo-\ko$ plane for various $N$'s indicated in the plot. Shown is
also the variation of $\vrm$ in grey along the lines.}\label{fig3}
\end{figure}

To localize the position of the inflection point, we impose
\beq \Vhi'(z)=\Vhi''(z)=0~~\mbox{for}~~\vrm<z<1, \label{ifp}\eeq
where prime stands for derivation w.r.t $z$. These conditions can
be translated as follows
\beqs\begin{align} &
\sqrt{3N}\Om+N(\Om+\om\Om')=2\Om U\frac{U+\om U'}{U^2-12\om^2}; \label{con1}\\
&N\Om\om\lf 4\Om'(\nnp-1)U^2+\om\lf12\om\lf\sqrt{3N}+N^2\rg+2UU'\rg\rg\nonumber\\
%
&+\Om^2\Big( U^2N_1+\om\big(4(1-\nnp)
U^{2\prime}-\om\lf12N_2-(U^2)'')\rg\big)\Big)\nonumber\\
& =(12\om^2-U^2)\lf\Om''\om +N(1+N)\Om'\om^2\rg,
\label{con2}\end{align} \eeqs  where we define the quantities
\beq \begin{aligned}
N_1&=6-5\sqrt{3N}-2N+\sqrt{3N^3}+N^2;\\N_2&=2N-\sqrt{3N}+2\sqrt{3N^3}+N^2.\end{aligned}
\label{Ns}\eeq
Note that the conditions above are independent from the parameters
$m$ and $\cm$ of $W_\Lambda$. Solving the conditions above w.r.t
$k$ and $\sg$, for every selected $\vrm$ and $N$, we can specify
$k=\ko$, for which we obtain an inflection point, and its position
at $\sg=\sgo$. The output of this procedure is given in
\Fref{fig3} where we plot $(\ko,\sgo)$ for $N=4, 10$ and $30$
(dashed, solid and dot-dashed line respectively). Along each line
we show the variation of $\vrm$ in grey. We observe that the
maximal allowed $\vrm$ values decrease as $N$ increases whereas
for $\vrm<0.01$ the $(\ko,\sgo)$'s remain almost constant. We also
remark that $\sgo>\vrm$ and it remains subplanckian whereas the
required $\ko<1$ is quite natural, although it has to be
determined using a precision up to six digits to be reliable.

\subsection{\sc\small\sffamily Inflation Analysis}\label{sec42}

Once we determine the $(\ko,\sgo)$ values, we can investigate the
realization of IPI which is delimited by the condition
\cite{review}
\beqs\beq{\ftn\sf max}\{\ep(\se),|\eta(\se)|\}\leq
1,\label{srcon}\eeq
where the slow-roll parameters $\ep$ and $\eta$ read
\beq\label{sr}\epsilon= \left({\Ve_{{\rm I},\se}/\sqrt{2}\Ve_{{\rm
I}}}\right)^2~~\mbox{and}~~\eta={\Ve_{{\rm I},\se\se}/\Ve_{\rm I}}
\eeq\eeqs
and can be derived by employing $\Vhi$ in \Eref{Vde} for $\th=0$
and $J$ in \Eref{jdef}, without express explicitly $\Vhi$ in terms
of $\se$. Due to the complicate form of $\Vhi$, the analytic
approach to the inflationary dynamics is not doable. Some progress
can be made, if we use as input for our analytic treatment the
numerical expansions of $\Vhi$ and $J$ about $z=\sgo$,
\beq \Vhi\simeq\vo+\va\dz+\vb\dz^2+\vc\dz^3~~\mbox{and}~~
J\simeq\jo, \label{vappr}\eeq
where $\dz=z-\sgo$, $\vo=\Vhi(\sgo)$ and $\jo=J(\sgo)$. Since the
observationally relevant part of IPI takes place quite close to
$\sgo$, the approximation above is quite accurate. This is
confirmed in \Fref{fig2} where the dotted line, obtained by
\Eref{vappr}, is plotted against the exact result of $\Vhi$ for
the inputs in \Tref{tab}. The relevant expansion coefficients are
listed there. Thanks to the conditions in \Eref{ifp},
$\va=\Vhi'(\sgo)$ and $\vb=\Vhi''(\sgo)/2$ are quite suppressed
compared to $\vo$ and $\vc$ and so we may neglect terms with
$\va^2, \vb^2$ and $\va\vb$ below. Namely, inserting \Eref{vappr}
into \Eref{sr} we arrive at the following results
\beq\label{sr1}
\sqrt{\eph}\simeq\frac{\va+\dz(2\vb+3\dz\vc)}{\sqrt{2}\jo\vo}~~\mbox{and}~~
\eta\simeq \frac{2(\vb+3\dz\vc)}{\jo^2\vo},\eeq
from which we can verify that \Eref{srcon} is saturated for
$\dz=\dzf$, found from the condition
\beq
\eta\lf\dzf\rg\simeq1~~\Rightarrow~~\dzf\simeq-(\jo^2\vo+2\vb)/6\vc.
\label{sgf}\eeq
Given that $\jo^2\vo\gg\vb$, we expect $\dzf<0$ or $\sgf<\sgo$.

The number of e-foldings $\Ns$ that the scale $\ks=0.05/{\rm Mpc}$
experiences during IPI and the amplitude $\As$ of the power
spectrum of the curvature perturbations generated by $\sg$ can be
computed using the standard formulae
\begin{equation}
\label{Nhi} \Ns=\int_{\sef}^{\sex} d\se\frac{\Vhi}{\Ve_{{\rm
I},\se}}~~~\mbox{and}~~~\As= \left.\frac{1}{12\pi^2}\frac{\Ve_{\rm
I}^{3}}{\Ve^2_{{\rm I},\se}}\right|_{\se=\sex},\eeq
where $\sgx~[\sex]$ is the value of $\sg~[\se]$ when $\ks$ crosses
the inflationary horizon. From the leftmost relation we find
\beq  \label{Nsa}
\Ns=(\fns-\fnf)/\pn~~\mbox{where}~~\pn={\sqrt{3\va\vc}}/{\jo^2\vo}\eeq
and $\fns=\fn(\dzx)$ and $\fnf=\fn(\dzf)$ with
\beq \fn(\sg)=\arctan\lf{\vb+3\sg\vc}/{\sqrt{3\va\vc}}\rg.
\label{fn}\eeq
Since $\dzf$ turns out to be close to $\dzx$ -- as depicted in
\Fref{fig2} -- both contributions in \Eref{Nsa} are important.
Solving it w.r.t $\dzx$ we obtain
\beq \dzx\simeq-\frac{\vb}{3\vc}+\sqrt{\frac{\va}{3\vc}}
\tan\lf\frac{\sqrt{3}\Ns}{\jo^2\vo}+\fnf\rg,\label{sgx}\eeq
from which we can deduce that $\sgx<\sgo$ -- see inset of
\Fref{fig2}. Plugging it into the rightmost equation in
\Eref{Nhi}, we obtain
\beq
\As^{1/2}\simeq\frac{\jo\vo^{3/2}}{2\sqrt{3}\pi\va}\cos^{2}\lf\pn\Ns+\fnf\rg.\label{Asa}\eeq

The remaining inflationary observables are found from
\beqs\bea \label{ns} && \ns=\: 1-6\epsilon_\star\ +\
2\eta_\star,~~r=16\epsilon_\star, \\ \label{as} && \as
=\:2\left(4\eta_\star^2-(\ns-1)^2\right)/3-2\xi_\star, \eea\eeqs
where $\xi={\Ve_{{\rm I},\widehat\sg} \Ve_{{\rm
I},\widehat\sg\widehat\sg\widehat\sg}/\Ve^2_{\rm I}}$ and the
variables with subscript $\star$ are evaluated at $\sg=\sgx$.
Inserting $\dzx$ from \Eref{sgx} into \Eref{sr} and then into
equations above we obtain
\beqs\bea\ns &\simeq& 1 +4\pn\tan\lf\pn\Ns+\fnf\rg,\label{nsa}\\
r &\simeq&8\va^2\cos^{-4}\lf\pn\Ns+\fnf\rg/\jo^2\vo^2, \label{ra} \\
\as &\simeq&-4\pn\cos^{-2}\lf\pn\Ns+\fnf\rg.\label{asa}\eea\eeqs
For the inputs of \Tref{tab}, the results of our semianalytic
approach are displayed in curly brackets and compared with those
obtained using the pure numerical program. The proximity of both
results is certainly impressive.

From \Tref{tab} we may also infer that the semiclassical
approximation, used in our analysis, is perfectly valid since
$\Vs^{1/4}\ll\mP$. Moreover, the $\th=0$ direction is well
stabilized and does not contribute to the curvature perturbation,
since for the relevant effective mass $m_{\th\rm I}$ we find
$m^2_{\th\rm I}>0$ for $N>3$ and $m_{\th\rm I\star}/\Hs>1$ where
$\Hhi=(\Vhi/3)^{1/2}$. We also checked that the one-loop radiative
corrections, $\dV$, to $\Vhi$ induced by $m_{\th\rm I}$ let intact
our inflationary outputs, provided that we take for the
renormalization-group mass scale $Q=m_{\th\rm I\star}$. Since $Q$
is close to $\Vs^{1/4}$ we do not expect sizable running of the
quantities measured at $Q$  -- cf.~\cref{jhep}.


\subsection{\sc\small\sffamily  Sgoldstino Decay}\label{sec43}

Soon after the end of IPI, the (canonically normalized) sgoldstino
\beq\dzvh=\vev{J}\dzv~~\mbox{with}~~\dzv=z-\vrm~~\mbox{and}~~\vev{J}=\sqrt{\frac{N}{2}}\frac{1}{\vev{\om}}
\label{dphi}\eeq
settles into a phase of damped oscillations abound the minimum in
\Eref{vcc} reheating the universe at a temperature \cite{rh}
\beqs\beq \label{Trh} \Trh= \left({72/5\pi^2g_{\rm
rh*}}\right)^{1/4}\Gsn^{1/2}\mP^{1/2},\eeq where $g_{\rm
rh*}=106.75$ counts the effective number of the relativistic
degrees of freedom at $\Trh$. The total decay width, $\Gsn$, of
$\dzvh$ predominantly includes the contributions
\beq\Gsn\simeq\Ggr+\Gth+\Gh,\label{Gol}\eeq\eeqs
where the individual decay widths -- which stem from SUGRA-induced
interactions \cite{full,baerh,antrh,nsrh} -- are found to be
%
%
\beq
(\Ggr,\Gth,\Gh)\simeq\lf\frac{\vev{\om}^{-\sqrt{3N}}\msn^5}{96\pi
m^2\mP^2},~~\frac{\msn^3}{16N\pi
\vrm\mP},\frac{N\what\mu^2\msn}{16\pi\mP^2}\rg.\label{Gs}\eeq
They express decay of $\dzvh$ into gravitinos, pseudo-sgoldstinos
and higgsinos via the $\mu$ term respectively. Thanks to the
appearance of $N$ in $\Gh$, it is rather enhanced for large $N$'s.


\subsection{\sc\small\sffamily Parameter Space}\label{sec44}


\begin{figure}[!t]\vspace*{-.25in}\centering
\includegraphics[width=60mm,angle=-90]{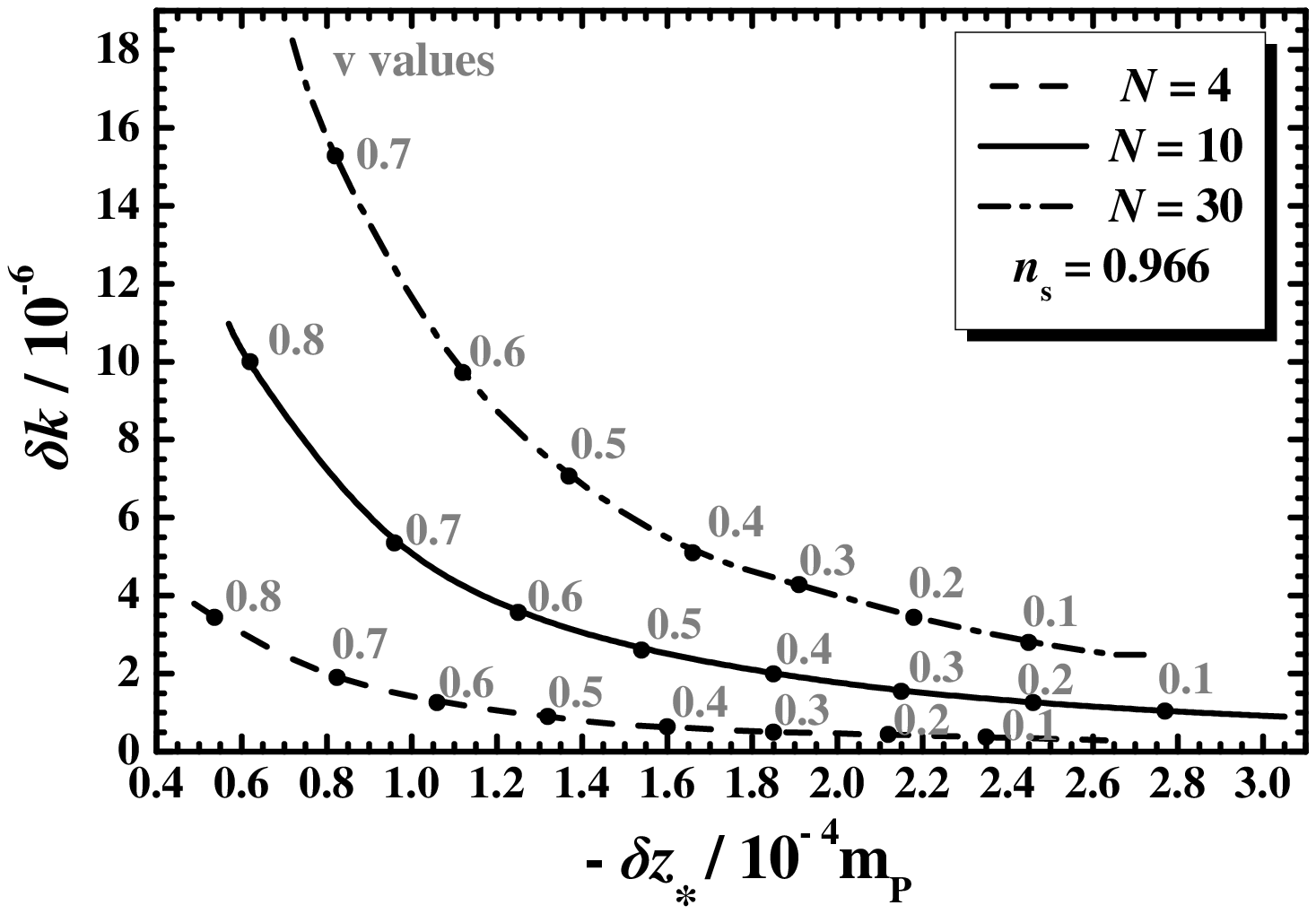}
\caption{\sl Allowed curves in the $(-\dzx)-\dk$ plane for
$\ns\simeq0.966$ and various $N$'s indicated in the plot. Shown is
also the variation of $\vrm$ in grey along the lines.
}\label{fig4}
\end{figure}


In order to delineate the available parameter space of the model,
we confront the quantities in \Eref{Nhi} with the observational
requirements \cite{plcp}
\beq\label{Ntot}
\Ns\simeq61+\ln\lf\pi\vo\Trh^2\rg^{1/6}~~\mbox{and}~~\As\simeq2.1052\cdot10^{-9},\eeq
where we assume that IPI is followed in turn by an oscillatory
phase, with mean equation-of-state parameter $w_{\rm rh}\simeq0$,
radiation and matter domination. The remaining observables must be
in agreement with the fitting of the \plk\ TT, TE,
EE+lowE+lensing, {\sffamily\ftn BK15} (from {\sc Bicep2}/{\it
Keck} Array) \cite{bk15} and BAO data \cite{plin} with the
$\Lambda$CDM$+r+\as$ model which requires
\begin{equation}  \label{nswmap}
\ns=0.9658\pm0.008~~\mbox{and}~~\as=-0.0066\pm0.014
\end{equation}
at 95$\%$ confidence level. Recall that the upper bound on $r$ is
irrelevant in our case since $r$ is negligible -- see
\Sref{sec42}.

In our numerical program for any selected $N$ and $\vrm$ -- see
\Eref{Vde} -- we compute $(\ko,\sgo)$ solving numerically
\Eref{ifp}. Then enforcing \Eref{Ntot} we can restrict $\dk=k-\ko$
and $m$ whereas the leftmost observable in \Eref{nswmap}
determines $\dzx$. From \Eref{as} the model's predictions
regarding $\as$ and $r$ can be computed. Increasing $\dk$ allows
us to increase the slope of the plateau around $\sgo$ decreasing,
thereby, $\Ns$. Note that the $\dk$ does not appear in the
formulae in \Sref{sec42}, since the relevant information is
encoded in \Eref{vappr}.

\begin{figure}[!t]\vspace*{-.25in}\centering
\includegraphics[width=60mm,angle=-90]{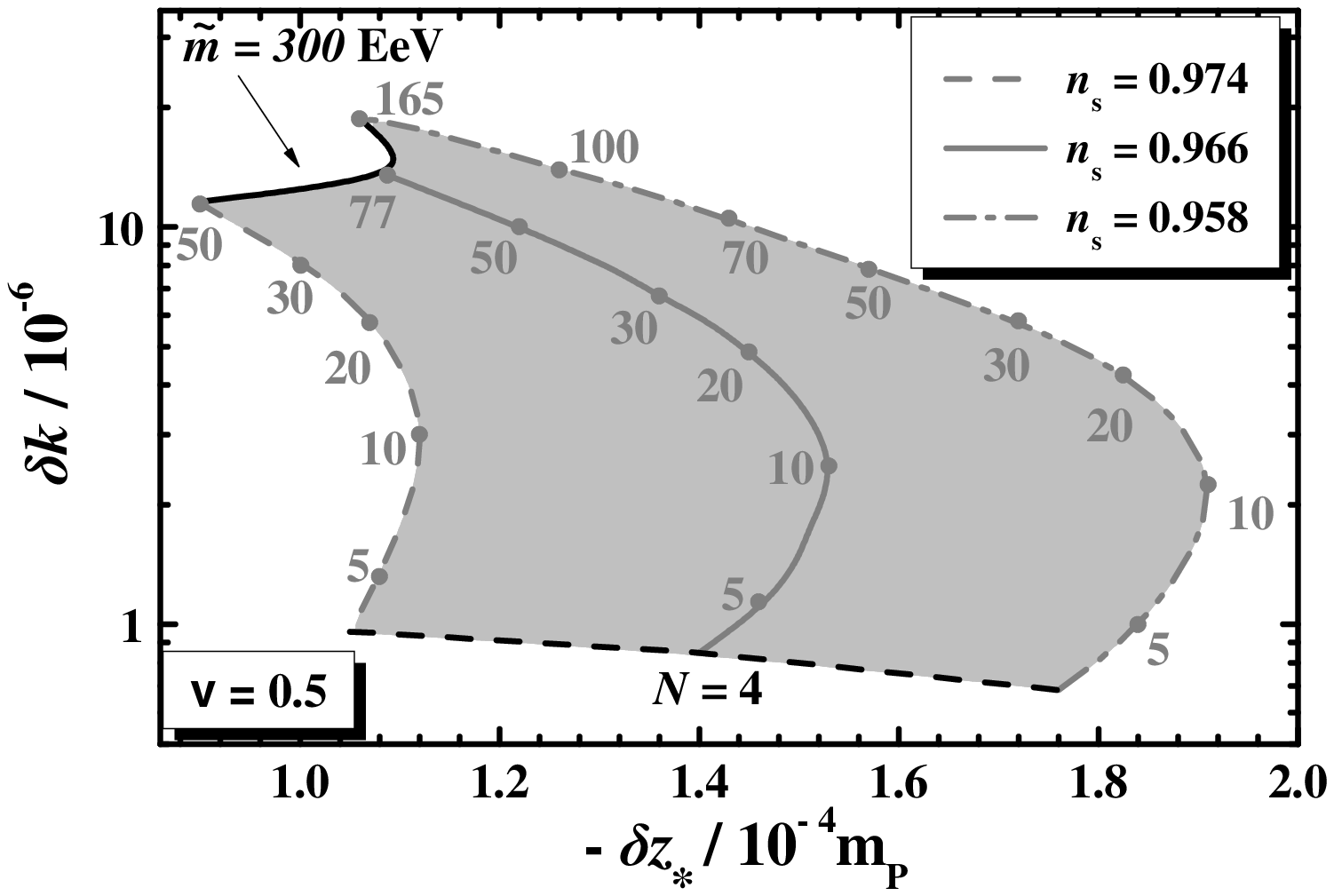}
\caption{\sl Allowed (shaded) region in the $(-\dzx)-\dk$ plane
for $\vrm=0.5$. The constraint fulfilled along each line and the
variation of $N$ are also shown in black and grey
respectively.}\label{fig5}
\end{figure}

The outputs of our numerical investigation can be presented in the
$(-\dzx)-\dk$ plane as in Figs.~\ref{fig4} and \ref{fig5}. In the
first one we fix $N$ to three representative values $4, 10$ and
$30$ and display the allowed curves (dot-dashed, solid and dashed
lines respectively) taking the central $\ns$ value in
\Eref{nswmap}. The variation of $\vrm$ along each line is
displayed in gray.  On the other hand, in \Fref{fig5} we set
$\vrm=0.5$ and identify the allowed (shaded) region allowing $\ns$
to vary in the margin of \Eref{nswmap}. The variation of $N$ is
shown along each line. The allowed region is bounded by {\sf\ftn
(i)} the solid black line, which corresponds to the upper bound in
\Eref{highb}, {\sf\ftn (ii)} the dashed black line which
originates from the lower bound on $N$ derived in \Sref{sec2} and
{\sf\ftn (iii)} the dot-dashed and dashed gray lines along which
the lower and upper bounds on $\ns$ in \Eref{nswmap} are saturated
respectively. We remark that increasing $|\dzx|$, decreases $\ns$
with fixed $\dk$. From both figures we deduce that the achievement
of observationally acceptable IPI requires a tuning of the order
of $10^{-6}$ which is somehow ameliorated as $N$ increases. This
tuning though is milder than that needed within the conventional
MSSM \cite{aterm}.

Fixing $\ns\simeq0.966$, we obtain the gray solid line in
\Fref{fig5}, along which we obtain the mass spectrum shown in
\Fref{fig6} as a function of  $m$. Namely, we depict $\mgr$, $\mz$
and $\mth$ -- calculated by \eqss{mgr}{mt}{mbt} respectively -- as
functions of $m$ employing solid, dashed and dot-dashed lines
correspondingly. Shown is also the variation of $N$ in grey along
the solid line. In all, we have
\beq\label{res2a} 2.3\lesssim
\frac{\msn}{100~\EeV}\lesssim4.4~~\mbox{and}~~8.9\lesssim
\frac{\mth}{10~\EeV}\lesssim59,\eeq
where the lower bound on $\mth$ coincides with that on
$\mgr=\mss$. Note that the $\dzvh$ decay channel into $\theta$'s
is kinematically blocked for $N\gtrsim20$. We also find that for
$N\lesssim10$, $\Gsn\simeq\Ggr+\Gh$ whereas for larger $N$'s
$\Gsn\simeq\Gh$ and so \Eref{Trh} yields $\Trh\simeq(4-20)~\PeV$
resulting to $\Ns\simeq(45.5-46.7)$. The obtained
$\as\simeq-(3.1-3.2)\cdot10^{-3}$ might be detectable in future
\cite{asdrees}.


\begin{figure}[!t]\vspace*{-.25in}\centering
\includegraphics[width=60mm,angle=-90]{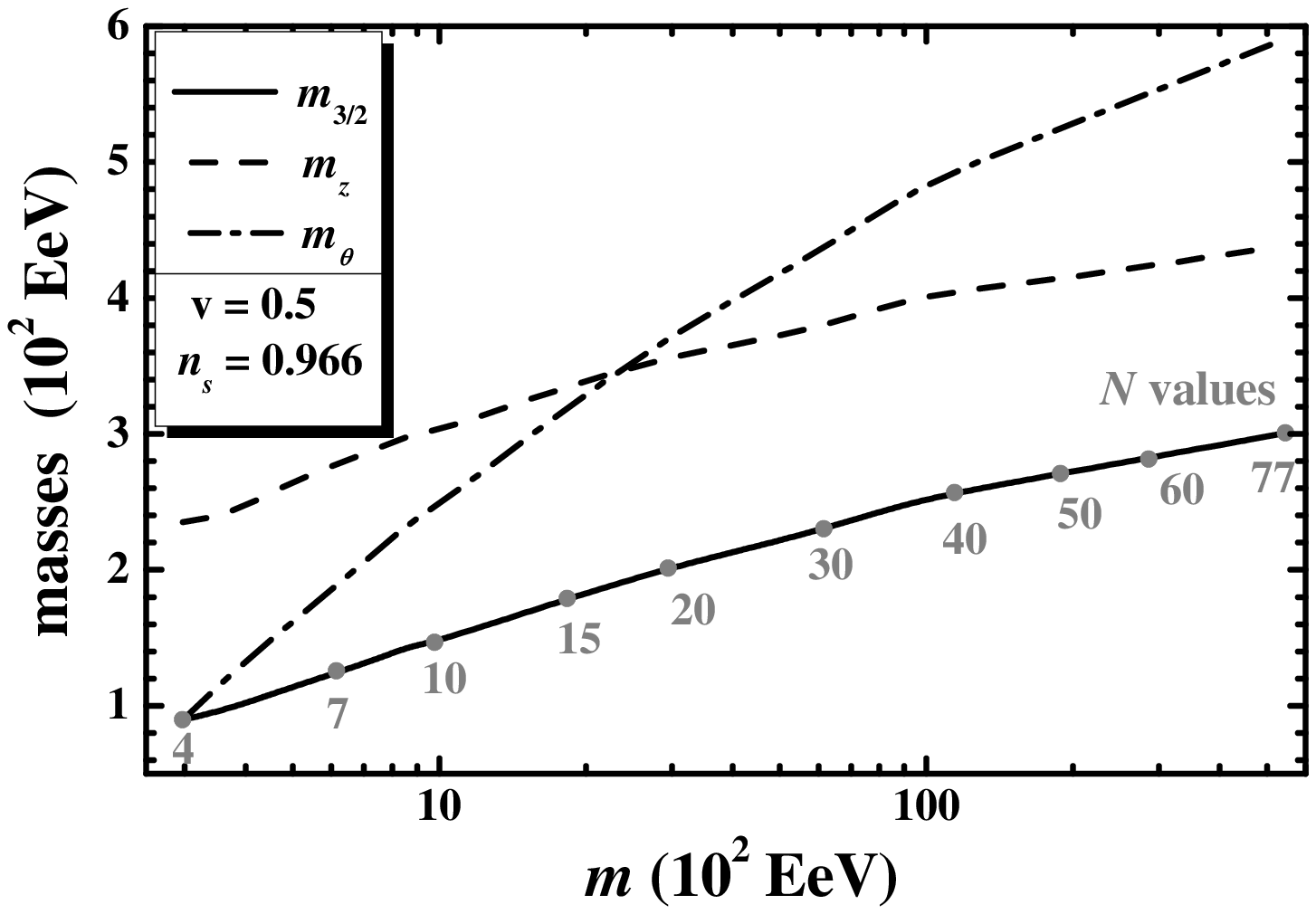}
\caption{\sl Allowed values of $\mgr$, $\mz$ and $\mth$ (solid,
dashed and dot-dashed line respectively) versus $m$ for
$\ns\simeq0.966$ and $\vrm=0.5$. Shown is also the variation of
$N$ in grey along the solid line.}\label{fig6}
\end{figure}

Throughout our investigation, we assumed that the slow-roll
approximation offers a reliable description of IPI. This is a
reasonable assumption since the observationally relevant part of
IPI takes place for $\sgx<\sgo$ -- see e.g. inset of \Fref{fig2}.
However, we do not address in this letter the question of how
$\sg$ reaches $\sgo$, that is, the problem of the initial
conditions for inflation. Since $V_{\rm I}$ is extremely flat
close to $\sgo$, there is the danger that the system temporarily
undergoes a period of the so-called ultra-slow-roll evolution
\cite{bh,usrdim,wands}, where the gradient of $V_{\rm I}$ may be
neglected rather than the acceleration of $\sg$ in the
Klein-Gordon equation. However, this danger can by averted
\cite{usrdim}, if we assume that the $\sg$ lies initially near
$\sgo$ with a small enough kinetic energy density which is at most
the one corresponding to the slow-roll phase,
\beqs\beq\rho_{\rm K}^{\rm max}\simeq\ep(\sgo)V_{\rm
I}(\sgo)/3.\eeq
Under such a condition, the slow roll starts immediately and all
our findings are perfectly valid. E.g., for the range of the
parameters in \Fref{fig6}, we find
\beq 1.4\lesssim\lf\rho_{\rm K}^{\rm
max}\rg^{1/4}/10^3~\EeV\lesssim 3.2,\eeq\eeqs
where the upper/lower bound corresponds to the respective $N$
bounds in \Fref{fig6} and turn out to be two orders of magnitude
less than $\Vhi(\sgo)^{1/4}$. Moreover, as shown for similar
models \cite{wands}, we can always find suitable initial
conditions in the phase space of the system so as slow-roll IPI to
take place. Since $\ep(\sg)$ in a such case is a smooth increasing
function without spikes, no production of black holes
\cite{bh,usrdim} occurs. \vspace{-5mm}

\section{\bfseries\scshape Conclusions} \label{con}

Taking advantage from the self-stabilized no-scale models of SUSY
breaking, established in \cref{ns89,de}, we proposed a variant
which incorporates IPI -- i.e., inflection-point inflation --
realized by the sgoldstino. The inflationary model results to an
adjustable $\ns$, a small $r$ and a sizable $\as$ of the order of
$-10^{-3}$. Linking our model to MSSM we showed that SUSY may be
broken at a dS vacuum, providing the correct DE density parameter
-- at the cost of a fine-tuned parameter -- and a SUSY mass scale
$\mss\sim 100~\EeV$ which is consistent with the Higgs boson mass
measured at LHC. Needless to say, the stability of the electroweak
vacuum  up to the Planck scale is automatically assured within our
framework.

It would be interesting to investigate if the intermediate-scale
lightest neutralino with mass $M_1$ in the interval
$\Trh<M_1<T_{\rm max}$ is a good cold dark matter candidate
adapting the non-equilibrium production \cite{kolb}. Here, $T_{\rm
max}$ is the maximal temperature during reheating and may be as
high as $10^3\Trh$ \cite{kolb}. Also, baryogenesis via non-thermal
leptogensis \cite{ntlepto} may be activated even without direct
coupling of the sgoldstino to right-handed neutrinos. Our model,
most probably, does not belong to the string swampland \cite{vafa}
but it is amenable to modifications \cite{sevilla} which may
render it more friendly with the string ultraviolet completions.


\paragraph*{\small\bfseries\scshape Acknowledgments} {\small I would like to thank Mar Bastero-Gil for
an interesting discussion. This research work was supported by the
Hellenic Foundation for Research and Innovation (H.F.R.I.) under
the ``First Call for H.F.R.I. Research Projects to support Faculty
members and Researchers and the procurement of high-cost research
equipment grant'' (Project Number: 2251).}


\def\ijmp#1#2#3{{\sl Int. Jour. Mod. Phys.}
{\bf #1},~#3~(#2)}
\def\plb#1#2#3{{\sl Phys. Lett. B }{\bf #1}, #3 (#2)}
\def\prl#1#2#3{{\sl Phys. Rev. Lett.}
{\bf #1},~#3~(#2)}
\def\rmpp#1#2#3{{Rev. Mod. Phys.}
{\bf #1},~#3~(#2)}
\def\prep#1#2#3{{\sl Phys. Rep. }{\bf #1}, #3 (#2)}
\def\prd#1#2#3{{\sl Phys. Rev. D }{\bf #1}, #3 (#2)}
\def\npb#1#2#3{{\sl Nucl. Phys. }{\bf B#1}, #3 (#2)}
\def\npps#1#2#3{{Nucl. Phys. B (Proc. Sup.)}
{\bf #1},~#3~(#2)}
\def\mpl#1#2#3{{Mod. Phys. Lett.}
{\bf #1},~#3~(#2)}
\def\jetp#1#2#3{{JETP Lett. }{\bf #1}, #3 (#2)}
\def\app#1#2#3{{Acta Phys. Polon.}
{\bf #1},~#3~(#2)}
\def\ptp#1#2#3{{Prog. Theor. Phys.}
{\bf #1},~#3~(#2)}
\def\n#1#2#3{{Nature }{\bf #1},~#3~(#2)}
\def\apj#1#2#3{{Astrophys. J.}
{\bf #1},~#3~(#2)}
\def\mnras#1#2#3{{MNRAS }{\bf #1},~#3~(#2)}
\def\grg#1#2#3{{Gen. Rel. Grav.}
{\bf #1},~#3~(#2)}
\def\s#1#2#3{{Science }{\bf #1},~#3~(#2)}
\def\ibid#1#2#3{{\it ibid. }{\bf #1},~#3~(#2)}
\def\cpc#1#2#3{{Comput. Phys. Commun.}
{\bf #1},~#3~(#2)}
\def\astp#1#2#3{{Astropart. Phys.}
{\bf #1},~#3~(#2)}
\def\epjc#1#2#3{{Eur. Phys. J. C}
{\bf #1},~#3~(#2)}
\def\jhep#1#2#3{{\sl J. High Energy Phys.}
{\bf #1}, #3 (#2)}
\newcommand\jcap[3]{{\sl J.\ Cosmol.\ Astropart.\ Phys.\ }{\bf #1}, #3 (#2)}
\newcommand\njp[3]{{\sl New.\ J.\ Phys.\ }{\bf #1}, #3 (#2)}
\def\prdn#1#2#3#4{{\sl Phys. Rev. D }{\bf #1}, no. #4, #3 (#2)}
\def\jcapn#1#2#3#4{{\sl J. Cosmol. Astropart.
Phys. }{\bf #1}, no. #4, #3 (#2)}
\def\epjcn#1#2#3#4{{\sl Eur. Phys. J. C }{\bf #1}, no. #4, #3 (#2)}


\end{document}